\documentstyle[11pt,newpasp,twoside,epsf]{article}

\newcommand{\lya}{\hbox{Ly$\alpha$}}

\begin{document}

\title{Deep Ly$\alpha$ imaging of radio galaxy 1138$-$262 at redshift 2.2}
\author{J.~D. Kurk, H.~J.~A. R\"ottgering, L. Pentericci, G.~K. Miley}
\affil{Sterrewacht Leiden, P.O. Box 9513, 2300 RA Leiden, The Netherlands}

\begin{abstract}
Observations of the powerful radio galaxy 1138$-$262 at $z=2.2$
suggest that this galaxy is a massive galaxy in the center of a
forming cluster. We have imaged the field of 1138$-$262 with the Very
Large Telescope\footnote{Based on observations collected as part of
programme 63.O-0477 at the European Southern Observatory, Paranal,
Chile.} in a narrow band encompassing the redshifted \lya\
emission. We detect 34 probable line emitters in the $6' \times 6'$
field. These are candidate \lya\ emitting galaxies in the supposed
cluster. The observations also reveal that the radio galaxy is
enclosed by a very extended \lya\ halo ($\sim$ 160 kpc).
\end{abstract}


\section{Introduction}
Observations of clusters at high redshift ($z > 2$) can directly
constrain cosmological models, but searches based on colors or narrow
band emission have not discovered more than a handful of presumed
cluster galaxies (Le F\`evre et al.\ 1996; Cowie \& Hu 1998). There are
several indications that powerful radio galaxies at high redshift
(HzRGs) are located at the centers of forming clusters. The powerful
radio galaxy 1138$-$262 has extensively been studied and there is
strong evidence that it is a forming brightest cluster galaxy in a
(proto-)cluster (e.g.\ Pentericci et al.\ 1997).  The arguments
include (i) the very clumpy morphology of 1138$-$262 as observed by
the HST (Pentericci et al.\ 1998), reminiscent of a merging system;
(ii) the extremely distorted radio morphology and the detection of the
largest radio rotation measures ($\sim$ 6200 rad m$^{-2}$) in a sample
of more than 70 HzRGs, indicating that 1138$-$262 is surrounded by a
hot and dense magnetized medium (Carilli et al.\ 1997); (iii) the
detection of extended X-ray emission around 1138$-$262 (Carilli et
al.\ 1998), indicating the presence of hot cluster gas.

\section{A cluster at redshift 2.2?}
With the aim of detecting \lya\ emitting cluster galaxies, the field
of 1138$-$262 was observed on April 12 and 13 1999 with FORS1 on the
VLT ANTU using a narrow band (65 \AA) covering the redshifted
Ly$\alpha$ (3814 \AA), and the broad B band which encompasses the
narrow band. The resulting \lya\ image shows a huge ($\sim$ 160
kpc)\footnote{We adopt a Hubble constant of $H_0$=50 km
s$^{-1}$Mpc$^{-1}$ and a deceleration parameter of $q_0$=0.5.} halo of
ionized hydrogen around the galaxy, which extends even further than
the radio emission.

From a combined \lya\ and B band image we have extracted $\sim$ 1600
sources with SExtractor (Bertin \& Arnouts, 1996), after a careful
consideration of the aperture size to be used for the
photometry. Objects, that are detected in the narrow band image at a
level 3$\sigma$ higher than expected from the broad band image, are
selected as candidate \lya\ emitters. Discarding 6 bright stars, we
detect 34 such objects in the 3$\times$3 Mpc$^2$ field with a range of
\lya\ fluxes from 0.1-5$\times 10^{-16}$ ergs s$^{-1}$
cm$^{-2}$. These are obvious candidates for being companion galaxies
in the cluster around 1138$-$262. Three of these candidates are shown
in Fig.\ 1. The next step will be to measure the redshifts of the
\lya\ emitters and subsequently determine the spatial correlation
function and the velocity dispersion, which together with the size of
the cluster will give a direct estimate of the total mass.

\begin{figure}
\plotfiddle{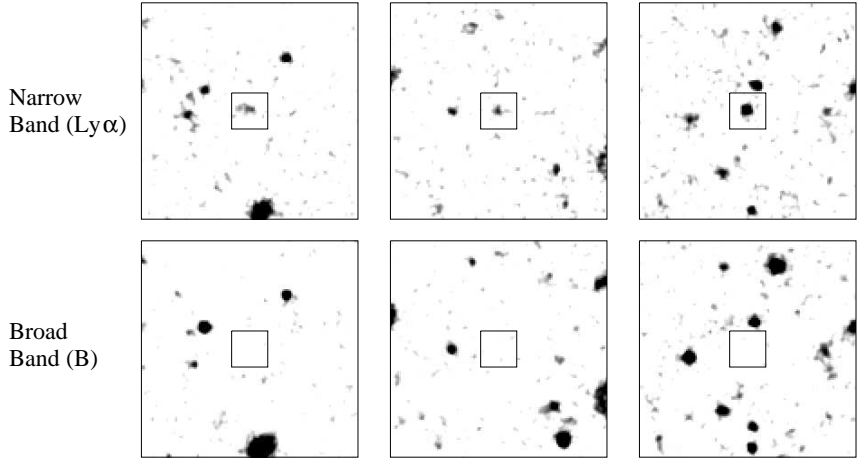}{5cm}{-90}{56}{56}{-232}{218}
\caption{Three probable cluster companions in NB(Ly$\alpha$) and BB(B)}
\end{figure}


\begin{references}
\reference Bertin, E. \& Arnouts S. 1996, \aaps, 117, 393
\reference Carilli, C. L., R\"ottgering, H. J. A., van Ojik, R., Miley,
G. K., \& van Breugel, W. J. M. 1997, \apjs, 109, 1
\reference Carilli, C .L., Harris, D. E., Pentericci, L., R\"ottgering,
H. J. A., Miley, G. K., \& Bremer, M. N. 1998, \apj, 496, L57
\reference Cowie, L. L. \& Hu, E. M. 1998, \aj, 115, 1319
\reference Le F\`evre, O., Deltorn, J. M., Crampton, D., Dickinson, M. 1996,
\apj, 471, L11
\reference Pentericci, L., R\"ottgering, H. J. A., Miley, C. L., \&
McCarthy, P. 1997, \aap, 326, 580
\reference Pentericci, L., R\"ottgering, H. J. A., Miley, G. K., Spinrad,
H., McCarthy, P. J., van Breugel, W. J. M., \& Macchetto, F. 1998, ApJ, 504,
139
\end{references}
\end{document}